\documentclass{PoS}

\title{The Electroweak Model based on the Nonlinearly realized Gauge Group.\\
\Large Theoretical Foundations and Phenomenological Prospects.}

\ShortTitle{The Nonlinearly Realized Electroweak Model}

\author{Daniele Bettinelli\\
        Albert-Ludwigs Universit\"at Freiburg\\
        E-mail: \email{daniele.bettinelli@physik.uni-freiburg.de}}

\author{Ruggero Ferrari\\
        CTP-MIT, Cambridge, MA\\
        and  Universit\`a degli Studi di Milano \& INFN, Sez. di Milano\\
          E-mail: \email{ruggero.ferrari@mi.infn.it}}

\author{\speaker{Andrea Quadri}\\
        Albert-Ludwigs Universit\"at Freiburg\\
        and  Universit\`a degli Studi di Milano \& INFN, Sez. di Milano\\
        E-mail: \email{andrea.quadri@mi.infn.it}}

\abstract{A consistent strategy for the subtraction of the divergences in
the nonlinearly realized Electroweak Model in the loop
expansion is presented. No Higgs field enters
into the perturbative spectrum.
The local functional equation (LFE), encoding the invariance
of the SU(2) Haar measure under local left SU(2) transformations,
the Slavnov-Taylor identity, required in order to fulfill
physical unitarity, and the Landau gauge equation hold in the
nonlinearly realized theory.
The quantization is performed in the Landau gauge for the
sake of simplicity and elegance.
The constraints on the admissible  interactions
arising from the Weak Power-Counting (WPC) are discussed. 
The same symmetric pattern of the couplings
as in the Standard Model is shown to arise, as a consequence
of the defining functional identities and the WPC. However, two independent mass
invariants in the vector meson sector are possible, i.e.
no tree-level Weinberg relation holds between the
$Z$ and $W$ mass.
Majorana neutrino masses can be implemented 
in the nonlinearly realized Electroweak Model
in a way compatible with the WPC and all the symmetries
of the theory.
}

\FullConference{RADCOR 2009 - 9th International Symposium on Radiative Corrections (Applications of Quantum Field Theory to Phenomenology) ,\\
		 October 25 - 30 2009\\
		 Ascona, Switzerland}

\def\G{\Gamma}

\begin{document}

\section{Symmetric Subtraction of Nonlinearly Realized
Gauge Theories}

A  consistent strategy for the
subtraction of nonlinearly realized 
gauge theories order by order in the loop
expansion has been recently proposed
in \cite{Bettinelli:2009wu}-\cite{Bettinelli:2007tq}.
The discovery of the Local Functional Equation
\cite{Ferrari:2005ii}, encoding the invariance of the SU(2)
path-integral Haar measure under local
left SU(2) transformations, has provided a key  tool
in order to tame the divergences of this class
of theories. 
The LFE uniquely fixes the dependence of
the 1-PI amplitudes involving at least
one Goldstone field (descendant amplitudes)
in terms of 1-PI amplitudes with no external
Goldstone legs (ancestor amplitudes).
This establishes a very powerful hierarchy
among 1-PI Green functions. 
While there is an infinite number of divergent
descendant amplitudes already at one loop level,
only a finite number of ancestor amplitudes exists
order by order in the loop expansion
if the Weak Power-Counting (WPC) condition
is fulfilled \cite{Bettinelli:2008qn},
\cite{Bettinelli:2008ey}, \cite{Bettinelli:2007tq}, \cite{Ferrari:2005va}.
In addition to the LFE, the Slavnov-Taylor (ST)
identity must be imposed in order to fulfill
the requirement of Physical Unitarity \cite{Ferrari:2004pd}.
It should be noted that the ST identity does
not yield a hierarchy among 1-PI Green functions
\cite{Bettinelli:2007tq}.
Thus the LFE provides an essential tool in order
to carry out the consistent subtraction of 
nonlinearly realized gauge theories.

The WPC poses stringent constraints on the 
admissible terms in the tree-level vertex functional.
In order to work out these constraints 
a convenient strategy is 
first to perform an invertible
change of variables from the original ones to their
corresponding SU(2) gauge-invariant counterparts (bleached variables).
We describe the procedure for the nonlinearly
realized Electroweak theory.
The field content includes (leaving aside  the ghosts
and the Nakanishi-Lautrup fields) the $SU(2)_L$
connection $A_\mu = A_{a\mu} \frac{\tau_a}{2}$
($\tau_a,~ a=1,2,3$ are the Pauli matrices), the 
$U(1)$ connection $B_\mu$, the fermionic left
doublets collectively denoted by $L$ and the right singlets, i.e.
\begin{eqnarray}
& L \in\Biggl\{ \left(
\begin{array}{r} l^u_{Lj}\\
l^d_{Lj}
\end{array} \right), \left(
\begin{array}{r}q^u_{Lj}\\
V_{jk}q^d_{Lk}
\end{array} \right), \quad j,k=1,2,3\Biggr\}, & \nonumber \\
& R \in\Biggl\{ \left(
\begin{array}{r}l^u_{Rj} \\
l^d_{Rj}
\end{array} \right), \left(
\begin{array}{r} q^u_{Rj}\\
q^d_{Rj}
\end{array} \right), \quad j = 1,2,3\Biggr\}. &
\label{sec.2.1}
\end{eqnarray} 
In the above equation the quark fields
$(q^u_j, j=1,2,3) = (u,c,t)$ and
$(q^d_j, j=1,2,3) = (d,s,b)$ are taken
to be the mass eigenstates in the tree-level
lagrangian; $V_{jk}$ is the CKM matrix.
Similarly we use for the leptons the notation
$(l^u_j, j=1,2,3) = (\nu_e,\nu_\mu,\nu_\tau)$ and
$(l^d_j, j=1,2,3) = (e,\mu,\tau)$.
The single left doublets are denoted by $L^l_j$,
$j=1,2,3$ for the leptons, $L^q_j$, $j=1,2,3$ for the 
quarks.
Color indexes are not displayed.

One also introduces the $SU(2)$ matrix $\Omega$
\begin{eqnarray}
\Omega = \frac{1}{v} (\phi_0 + i \phi_a \tau_a) \, , ~~~
\Omega^\dagger \Omega = 1 \Rightarrow \phi_0^2 + \phi_a^2 = v^2 \, .
\label{sec.2.2}
\end{eqnarray}
The  mass scale $v$  gives $\phi$ the canonical dimension
at $D=4$. We fix the direction of 
Spontaneous Symmetry Breaking
by imposing the tree-level constraint
\begin{eqnarray}
\phi_0 = \sqrt{v^2 - \phi_a^2} \, . 
\label{sec.2.3}
\end{eqnarray}
%
%The condition $\langle \Omega \rangle = 1$ cannot be imposed
%at a generic order of perturbation theory.

\par
The $SU(2)$ flat connection is defined by
\begin{eqnarray}
F_\mu = i \Omega \partial_\mu \Omega^\dagger \, .
\label{sec.2.4}
\end{eqnarray}
The local $SU(2)_L$ transformations 
act on $\Omega$ and $L$ on the left,
$A_\mu$ and $F_\mu$ are 
$SU(2)_L$ connections while
$R$ and $B_\mu$ are invariant
under $SU(2)_L$.
%($g$ is the $SU(2)_L$ coupling constant)
%
%\begin{eqnarray}
%\begin{array}{ll}
% \Omega' = U \Omega \, , & B'_\mu = B_\mu \, , \\
% A'_\mu = U A_\mu U^\dagger + \frac{i}{g} U \partial_\mu U^\dagger \, ,
% & L' = U L \, , \\
% F'_\mu = U F_\mu U^\dagger + i U \partial_\mu U^\dagger \, , 
% & R'=R \, .
%\end{array}
%\label{sec.2.5}
%\end{eqnarray}
%
Under local $U(1)_R$ transformations one has
\begin{eqnarray}
\begin{array}{ll}
 \Omega' = \Omega V^\dagger \, , & B'_\mu = B_\mu + \frac{1}{g'} \partial_\mu \alpha \, , \\
 A'_\mu = A_\mu ,
 & L' = \exp ( i \frac{\alpha}{2} Y_L) L \, , \\
 F'_\mu = F_\mu + i \Omega V^\dagger \partial_\mu V \Omega\, , 
 & R'=\exp ( i \frac{\alpha}{2} (Y_L + \tau_3) ) R  \, .
\end{array}
\label{sec.2.6}
\end{eqnarray}
where $V(\alpha) = \exp(i \alpha \frac{\tau_3}{2})$.
The electric charge is defined according to the Gell-Mann-Nishijima relation
\begin{eqnarray}
Q = I_3 + Y \, ,
\label{sec.2.6.bis}
\end{eqnarray}
where the hypercharge operator $Y$ is the generator of the
$U(1)_R$ 
transformations (\ref{sec.2.6})
and $I_3 = \frac{\tau_3}{2}$ is the 
third component of the weak isospin.
The introduction of the matrix $\Omega$ allows to perform an invertible
change of variables from the original set of fields to a new set
of $SU(2)_L$-invariant ones (bleaching procedure). 
For that purpose we define
\begin{eqnarray}
&& w_\mu = w_{a\mu} \frac{\tau_a}{2} = g \Omega^\dagger A_\mu \Omega - g' B_\mu \frac{\tau_3}{2}
+ i \Omega^\dagger \partial_\mu \Omega \, , \nonumber \\
&& \tilde L = \Omega^\dagger L \, .
\label{sec.2.7}
\end{eqnarray}
Both $w_\mu$ and $\tilde L$ are $SU(2)_L$-invariant, while under
$U(1)_R$ they transform as
\begin{eqnarray}
w'_\mu = V w_\mu V^\dagger \, , ~~~~
\tilde L' = \exp (i \frac{\alpha}{2} (\tau_3 + Y_L) ) \tilde L \, .
\label{sec.2.8}
\end{eqnarray}
%
%I.e. the electric charge coincides with the hypercharge on the
%bleached fields, as it is apparent from the comparison of eqs.(\ref{sec.2.6}),
%(\ref{sec.2.6.bis}) and (\ref{sec.2.8}).

%%%%%%%%%%%%%%%%%%%%%%%%%%%%%

Since the bleached variables are SU(2)-invariant,
the  hypercharge generator coincides on them with
the electric charge. Therefore any electrically neutral local monomial
depending on the bleached variables and covariant
derivatives w.r.t. the U(1) gauge connection $B_\mu$
is allowed on symmetry grounds.

%The requirement of the validity of the WPC imposes 
%a set of constraints among these monomials.
%It turns out that  
%all the symmetric interactions between ancestor
%amplitudes present in the Standard Model are 
%recovered \cite{Bettinelli:2008qn}, \cite{Bettinelli:2008ey} by imposing the symmetries
%and the WPC (no anomalous couplings are allowed).
%On the other hand, two independent mass invariants
%are possible in the vector meson sector. Therefore
%the tree-level Weinberg relation between the $Z$ and
%$W$ mass does not hold in the nonlinearly realized
%theory.

\section{The Weak Power-Counting}

%Any Lorentz-invariant electrically neutral local monomial built out of 
%the components of $w_{a\mu},\tilde L, R$ and covariant
%derivatives w.r.t. $B_\mu$ is allowed on symmetry grounds.
We further impose the validity of the WPC condition, i.e.
the superficial degree of divergence of 
 any 1-PI graph ${\cal G}$ with $N_A$ gauge boson external legs
and $N_F$ fermionic legs must be bounded
by
\begin{eqnarray}
d({\cal G}) = (D -2 ) n + 2 - N_A - N_F \, ,
\label{wpc.1}
\end{eqnarray}
where $D$ is the space-time dimension and $n$ is the number of loops.
Several comments are in order here. First of all in $D=4$ we see that
the number of ancestor divergent amplitudes  compatible with
the bound (\ref{wpc.1}) increases with the loop order $n$.
Therefore we refer to formula (\ref{wpc.1}) as the {\em weak}
power-counting condition.
Moreover from eq.(\ref{wpc.1}) one also sees that the number
of divergent ancestor amplitudes is finite at every loop order.
We also notice that the UV dimension of the fermion fields
in one in the nonlinearly realized theory (instead of $3/2$
as in the linearly realized models).
The reason is that the gauge-invariant mass terms $l^d_{Rj} \tilde l^d_j$ 
generate upon expansion in powers of the Goldstone fields a quadrilinear
vertex of the form $l^d_{Rj} \phi^2 l^d_j$ which contains
two Goldstone fields. Therefore at one loop level there are
logarithmically divergent graphs with four fermionic external legs  
 \cite{Bettinelli:2008qn}, \cite{Bettinelli:2008ey}
which limit the UV dimension of massive fermions
to $1$.

Since the fermions have UV degree 1, massive
Majorana neutrinos are allowed in the nonlinear
theory. In fact the following invariant can be constructed
out of the first component of $\tilde L$, provided that
$\nu$ is Majorana
$$ m \overline{{\tilde \nu_L}} \tilde \nu_L.$$

Moreover two mass invariants in the vector
meson sector are obtained out of the charged combinations
\begin{eqnarray}
w^\pm_\mu = \frac{1}{\sqrt{2}} (w_{1\mu} \mp i w_{2\mu}) 
\label{sec.2.9}
\end{eqnarray}
and the neutral component $w_{3\mu}$. 
They can be parameterized  as
\begin{eqnarray}
M^2 \Big ( w^+ w^- + \frac{1}{2} w_3^2 \Big ) \, , ~~~~ \frac{M^2 \kappa}{2} w_3^2 \, .
\label{sec.2.10}
\end{eqnarray}
Two mass invariants are expected for
the vector mesons, as a consequence of the breaking of the
global $SU(2)_R$ invariance induced by the hypercharge.
We remark that in the SU(2) nonlinearly realized Yang-Mills theory
the WPC plus the gauge symmetry were compatible
with any bilinear term with no derivatives in the bleached 
gauge field. The unique diagonal St\"uckelberg mass
term was recovered by imposing an additional $SU(2)_R$
global symmetry.

\section{Gauge-fixing and External sources}
In order to preserve the LFE after gauge-fixing 
an external source $V_{a\mu}$ transforming
as a $SU(2)_L$ gauge connection is introduced.
The Landau gauge-fixing is implemented through
the $SU(2)_L$-covariant gauge-fixing function 
${\cal F}_a = D_\mu[V](A-V)^\mu_a$.
The antifields coupled to the nonlinear BRST variation
of the fields are introduced, as well as the
source of the nonlinear constraint $\phi_0$.
Moreover the source $V_{a\mu}$ is paired
with its BRST partner $\Theta_{a\mu}$ into
a BRST doublet \cite{Bettinelli:2008qn} and
\cite{Bettinelli:2007tq}.

The algebra of the operators coupled with these
external sources closes and consequently
the functional formulation of the relevant
identities can be established as discussed
in \cite{Bettinelli:2008qn} and \cite{Bettinelli:2008ey}.

The extension of the WPC to the external
sources is discussed in \cite{Bettinelli:2008qn}.

\section{The subtraction procedure}

The perturbative expansion is carried out order by order in the number
of loops. The LFE is solved by extending the bleaching technique
to the full set of ghost fields and external sources 
\cite{Bettinelli:2008qn} and \cite{Bettinelli:2008ey}.
The ST identities can be recursively studied order by order in the loop
expansion. 
The constraints on the divergent ancestor amplitudes
are derived by exploiting the nilpotency
of the linearized ST operator ${\cal S}_0$.
One has then to solve a cohomological problem in the space
of bleached variables of finite dimension, due to the WPC condition
\cite{Bettinelli:2007kc}.

Finite higher order symmetric renormalization, allowed by
the WPC and the symmetries of the theory, cannot be reinserted
back into the tree-level vertex functional without violating
either the symmetries or the WPC. 
This fact implies that they cannot be interpreted as
additional {\em bona fide} physical parameters \cite{Bettinelli:2007zn}
(unlike in the chiral effective field theories approach).
We adopt the following {\em Ansatz}: minimal subtraction of properly normalized  $n$-loop amplitudes
$$ \frac{1}{\Lambda^{D-4}} \G^{(n)}$$ 
around $D=4$ should be performed.
This subtraction procedure is symmetric \cite{Bettinelli:2008qn}, \cite{Bettinelli:2008ey} and \cite{Bettinelli:2007zn}.
In this scheme the $\gamma_5$
 problem is treated in a pragmatic approach. 
 The matrix $\gamma_5$ 
 is replaced by a new $\gamma_D$ which anti-commutes with 
every $\gamma_\mu$ . No statement is made on the analytical properties of the traces involving $\gamma_D$ . 
Since the theory is not anomalous such traces never meet poles in 
$D-4$ and therefore we 
can evaluate at the end the traces at $D = 4$.

In this subtraction scheme
the dependence on the scale $\Lambda$ cannot be removed
by a shift of the  tree-level parameters. Hence it must be considered
as an additional physical parameter setting the scale of the
radiative corrections.

\section{Conclusions}\label{sec.concl}

The electroweak model based on the nonlinearly realized
$SU(2) \otimes U(1)$ gauge group can be consistently 
defined in the perturbative loop-wise expansion.
In this formulation there is no Higgs in the perturbative series.

The present approach is based on the LFE and the WPC.
There is a unique classical action giving rise to Feynman rules
compatible with the WPC condition.
In particular the anomalous couplings, which would be otherwise allowed
on symmetry grounds, are excluded by the WPC. 
Two gauge bosons mass invariants
are compatible with the WPC and the symmetries. Thus the
tree-level Weinberg relation is not working in the nonlinear
framework.

The discovery of the LFE suggests a unique 
{\em Ansatz} for the subtraction
procedure which is symmetric, i.e. it respects 
all the identities of the theory.
A linear Ward identity exists for the electric charge
(despite the nonlinear realization of the gauge group).
The strategy does not alter the 
number of tree-level parameters apart from a common mass
scale of the radiative corrections. 

The theoretical and phenomenological consequences of this scenario
are rather intriguing. 
An Higgs boson could emerge as a non-perturbative 
mechanism, but then its physical parameters are not
constrained by the radiative corrections of the low energy
electroweak processes. Otherwise the energy scale
for the radiative corrections $\Lambda$ is a manifestation
of some other high-energy physics.

Many aspects remain to be further studied. We only mention some
of them here.
The issue of unitarity at large energy (violation of Froissart bound) 
at fixed order
in perturbation theory when the Higgs field is removed 
can provide additional insight in the role of the mass scale $\Lambda$
and in the transition to the symmetric phase
of the Yang-Mills theory. 
The electroweak model
based on the nonlinearly realized gauge group
 satisfies Physical Unitarity as a consequence
of the validity of the Slavnov-Taylor identity. Therefore 
violation of the Froissart bound can only occur in evaluating cross sections
at finite order in perturbation theory. This requires the evaluation 
of a scale at each order where unitarity at large energy is substantially violated.

The phenomenological implications of the nonlinear theory
in the electroweak precision fit have to be investigated.
The 't Hooft gauge derived for the nonlinearly realized
$SU(2)$ massive Yang-Mills theory should be extended to
the $SU(2) \otimes U(1)$ nonlinearly realized model
\cite{Bettinelli:2007eu}.

The evaluation of the radiative corrections
at one loop level is currently being investigated.
The phenomenological impact of the inclusion
of massive Majorana neutrinos should also
be addressed.

Finally the extension of the present approach to
larger gauge groups
(as in Grand-Unified models)  could help in
understanding  the nonlinearly realized spontaneous
symmetry breaking  mechanism (selection
of the identity as the preferred direction in the $SU(2)$ manifold)  and the 
associated appearance of two independent gauge boson mass invariants.

\end{document}